\documentstyle[12pt]{article}

\def\half{{\textstyle{1\over2}}}

\begin{document}

\thispagestyle{empty}
\begin{titlepage}

\bigskip
\hskip 3.7in{\vbox{\baselineskip12pt
}}

\bigskip\bigskip\bigskip\bigskip
\centerline{\large\bf Thermal Tachyons and the $g$-Theorem }
\bigskip\bigskip
\bigskip\bigskip
\centerline{\bf Shyamoli Chaudhuri
\footnote{shyamoli@thphysed.org} } \centerline{214 North Allegheny
St.} \centerline{Bellefonte, PA 16823}
\date{\today}

\bigskip\bigskip
\begin{abstract}
We give a pedagogical introduction to Affleck and Ludwig's g-theorem, 
distinguishing
its applications in field theory vs string theory. 
We clarify the recent proposal that the vacuum degeneracy $g$ of a
noncompact worldsheet sigma model with a continuous spectrum of
scaling dimensions is lowered under renormalization group flow
while preserving the central charge. As an illustration we argue
that the IR stable endpoint of the relevant flow of the worldsheet 
RG induced by a thermal
tachyon in the type II string
is the noncompact supersymmetric vacuum with lower $g$. Note
Added (Sep 2005).
\end{abstract}

\end{titlepage}

\vskip 0.1in Recently, a spate of papers have examined the issue
of closed string tachyon condensation within the context of
localized tachyon states in a noncompact worldsheet sigma model,
analogous to the much better understood case of open string
tachyon condensation. By noncompact is meant a worldsheet
conformal field theory with an operator spectrum characterized by
a continuum of scaling dimensions. In such cases, the assumptions
under which Zamolodchikov's $c$-theorem holds are no longer
satisfied \cite{polch,aps}, and it is interesting to ponder the
question of what quantity other than the effective central charge
is minimized under worldsheet renormalization group flow. HKMM
have answered this question with the conjecture that the
appropriate quantity in these cases is the vacuum degeneracy of
bosonic states in the worldsheet sigma model. In this paper, we
will clarify the APS and HKMM conjectures and offer a
nonperturbative illustration of their validity in a remarkably
straightforward example.

\vskip 0.1in
We begin with a summary of the situation in two-dimensional conformal field theory.
In the earliest formulation of a $g$-theorem \cite{al}, Affleck and Ludwig make three
separate observations that suggest the possibility of an, in general, non-integer vacuum
degeneracy, $g$, in a conformal field theory defined on a strip of length $l$ and
width, or inverse temperature, $b$. Periodic boundary conditions are assumed in the $b$
direction for bosonic fields. In the case of a nontrivial boundary condition along the
edge of length $l$, the following statements are known to hold true:

\vskip 0.1in
\noindent i) in the large $l$ limit, the high energy spectrum of the conformal
field theory is expected to approach a continuum with energy levels $E_n$ scaling
as $n/l$, and the asymptotic degeneracy takes the well-known form:
\begin{equation}
\rho(n) \to g \left ({{c}\over{6n^3}} \right )^{1/4} \exp \left [ 2 \pi (cn/6)^{1/2} \right ]
\quad ,
\label{eq:density}
\end{equation}
where $c$ is the central charge. Clearly, there is no a priori reason for the coefficient
$g$ to be an integer.

\vskip 0.1in
\noindent ii) As was first emphasized by Cardy, a conformally invariant boundary condition
corresponds to a boundary state and so, interchanging the role of the coordinates on
the strip, this implies the following factorization condition for the transfer matrix
$ Z_{AB}$ between boundary states $|A>$, $|B>$, in the limit $l$$>>$$b$:
\begin{equation}
 Z_{AB} = <A| \exp \left [ - l H \right ] |B> \to <A|0> e^{-l E_0} <0|B> \quad .
\label{factzn}
\end{equation}
Here $E_0$ is the vacuum energy for the Hamiltonian computed with periodic boundary
conditions on the interval of length $b$. We identify the vacuum degeneracy $g$ with
the product $g_A g_B$, where $g_A$$=$$<A|0>$, $g_B$$=$$<B|0>$.

\vskip 0.1in
\noindent iii) The free energy associated with the boundary degrees of freedom in the
two-dimensional field theory,
$F$$=$$-{{1}\over{b}} {\rm ln}  Z$,
will typically contain both a term independent of the interval-length, $ f_{0}$,
where $f_0$ is the non-universal free energy density whose definition is ultraviolet
regulator dependent, and a term inversely proportional to the length,
$-{{1}\over{b}} {\rm ln} g$.
The universal contribution to the free energy, $-{\rm ln} g$, also shows up as a universal
contribution to the entropy, giving rise to the term {\em boundary entropy}. It is this
universal contribution which is conjectured to decrease under a renormalization group
flow induced by a relevant operator living on the boundary \cite{al}. Moreover, since the
bulk degrees of freedom in the conformal field theory are unaltered, the effective central
charge is preserved \cite{al,kmm,hkmm,martinec}.

\vskip 0.1in
We will now clarify the corresponding statements in string theory. The one-loop
amplitude in either a closed, or open and closed, string theory is given by an integral
over conformal field theories on a strip. In the open string case, we assume periodic
boundary conditions on bosonic worldsheet fields in the direction parallel to the
boundaries of the cylinder, while nontrivial boundary conditions may be
imposed in the direction orthogonal to the boundaries. The one-loop string amplitude
is given by an integral over the cylinder length, $l$, from $0$ to $\infty$.
The factorization limit of the cylinder samples the lowest lying closed string modes
which dominate in the $l$$\to$$\infty$ limit. By the worldsheet UV-IR correspondence,
this is equivalent to the high energy asymptotic behavior of the open string mass spectrum.
The sum over one-loop connected vacuum graphs in string theory is simply the one-loop
spacetime effective potential \cite{poltorus}

\vskip 0.1in
Since the cylinder graph factorizes on two copies of the disc, the high energy
asymptotics of the open string spectrum can alternatively be obtained by
computation of the disc amplitude. In fact, this is the
route more commonly taken in the string theory literature \cite{tseyt,hkmm,martinec}.
Note, in particular, that since one-loop string amplitudes are normalizable \cite{poltorus},
the disc amplitudes obtained in the factorization limit are also unambiguously
normalizable. This enables a precise identification of the perturbed worldsheet path
integral with a perturbative contribution to the spacetime effective action. Given a
relevant perturbation of the worldsheet action by the boundary operator, $O_i$,
the resulting change in the tree-level spacetime effective action is expressed in
terms of the path integral over worldsheets with the topology of the disc:
\begin{equation}
\Gamma_{\rm ST} (\lambda_i) = \int_{\rm disc} [dX] e^{S_0} {\rm Tr}
       \exp \left [ - \int_{\partial} ds \lambda_i O_i \right ] \quad .
\label{eq:disc}
\end{equation}
RG flow by a relevant operator in a unitary conformal field theory
can, therefore, be seen to correspond to rolling down the
effective potential to a more stable extremum \cite{hkmm}:
\begin{equation}
{{\partial \Gamma_{\rm ST}}\over{\partial \lambda_i}} = \beta_i(\lambda)G_{ij} , \quad
{{\partial \Gamma_{\rm ST}}\over{\partial \mu}} = - \beta_i \beta_j G_{ij}  \quad ,
\label{eq:beta}
\end{equation}
where $G_{ij}$ is the Zamolodchikov metric in coupling space, $\beta_i$ is the
perturbative beta function for the coupling $\lambda_i$, and $\mu$ is the worldsheet
renormalization scale. Since the Zamolodchikov metric is positive definite,
worldsheet RG flow is always in the direction of decreasing spacetime effective action,
and we have the equivalence $g$$=$$\half \Gamma_{\rm ST}^2$.

\vskip 0.1in
Moving on to closed string theories, the one-loop amplitude is
given by the torus graph with two worldsheet moduli, $\tau_1$, $\tau_2$. It is well-known
that performing the integral over $\tau_1$ is equivalent to imposing the level matching
condition on the string mass spectrum. Let us assume periodic boundary conditions in the
$\sigma_1$ direction. The factorization limit of the torus amplitude samples the lowest lying
closed string modes which dominate in the $\tau_2$$\to$$\infty$ regime of moduli space.
By the property of modular invariance, we know that the behaviour of the integrand in
this regime can equivalently be mapped to that in the deep ultraviolet regime, i.e.,
in the vicinity of the origin in the $(\tau_1, \tau_2 )$ plane. Thus, analogous to
the open string case, the factorization limit of the torus samples the high energy
asymptotics of the closed string mass spectrum \cite{poltorus,hkmm}. Notice, however, that
there is no obvious correspondence in closed string theory between a change in the
spacetime effective action and the tree level closed string amplitude. A technical
reason for this is the non-normalizability of the sphere amplitude or, equivalently,
the fact that the normalizable one-loop closed string graph does not factorize on
tree-level graphs. We will return to this point later in the paper. HKMM \cite{hkmm} have
pointed out that in cases where RG flow is induced by a {\em localized} closed string state,
the bulk central charge of the worldsheet theory will remain unchanged. They further
conjecture that the quantity which decreases
along the direction of flow in these cases is the number of localized bosonic degrees of
freedom in the worldsheet conformal field theory. This factor appears in the high energy
density of states as a prefactor multiplying the central charge, and is therefore a direct
analog of Affleck and Ludwig's $g$. We will refine HKMM's conjecture below by considering
directly the free energy of the string theory, rather than the density of states of the
worldsheet conformal field theory.

\vskip 0.1in
At this point, we have summarized the equivalent of statements (i) and (ii) above for
string theory. We reiterate that Affleck and Cardy's observations apply to two-dimensional
field theories, and therefore hold for the {\em integrand} of the one-loop string path
integral. They should not be applied directly to the one-loop string amplitude: the physical
state spectrum of a string theory contains fewer states than in the worldsheet conformal
field theories.

\vskip 0.1in
The analog of Affleck and Ludwig's statement (iii) in string theory is equally subtle
but appears to have received less attention in the literature. AL separate the free
energy of the two-dimensional field theory into universal and nonuniversal terms. The
normalizable one-loop
string path integral computes the one-loop contribution to the spacetime effective
action functional, ${\cal W}$ $=$ ${\rm ln} {\cal Z}$ \cite{poltorus}.
Consider a target space with Euclidean time of interval size $\beta$ equal to
the inverse temperature.
Then the free energy of the string theory, ${\cal F} $$=$$-{{1}\over{\beta}} {\cal W}$,
can be interpreted, in two dimensional terms, as a distribution over the free energies of
two dimensional field theories on a strip of varying length and varying inverse temperature.
However, unlike what is stated in (iii), the entire expression for $\cal F$
is {\em universal}: upon using a scale invariant, zeta-function regulator \cite{poltorus}
as dictated by the requirement of manifest two-dimensional gauge invariance, the resulting
expression for the free energy is free of infrared regulator-dependent ambiguity.
We emphasize that $W$ is an intensive quantity from the viewpoint of both target space and
worldsheet--- it is dimensionless. Thus, there is no string theory analog of Affleck and
Ludwig's $f_0$. The free energy is normalizable at any fixed point. Moreover, the property
of universality of the free energy holds for both closed, and open and closed, string theories.

\vskip 0.1in
We will now illustrate these concepts with a new example of a localized relevant perturbation
in a closed
string theory that does not fall within the class of orbifold twisted sector perturbations
considered by HKMM. Since the operator is localized in the target space, the central
charge of the worldsheet sigma model is preserved under RG flow, and we can confirm that
$g_{\rm IR}$$<$$g_{\rm UV}$ by direct computation of the one-loop free energy at both
fixed points. Moreover, since we are able to write down the explicit form of the free
energy at intermediate points along the flow, our example is a nonperturbative illustration
of the validity of the $g$-theorem conjectured by HKMM.

\vskip 0.1in
We begin with the expression for the one-loop free energy of the type IIA string on the
noncompact target space $R^9$, with Euclideanized time, $X_0$, an interval, $S^1/Z_2$,
of length $\beta$, equal
to the inverse temperature \cite{fermionic}. The ultraviolet fixed point of the RG flow
is the tachyon-free thermal vacuum at the so-called minimum temperature,
$T^2_{\rm min}$$=$$ 1/2\pi^2 \alpha^{\prime }$. The marginally relevant perturbation we shall
consider is a coherent superposition of the operators carrying thermal momentum, where we
sum over all mode numbers, $n$$=$$\pm 1$, $\cdots$, $\pm \infty$. The vertex operator for the $n$th
operator in the superposition can be written in the ghost number zero picture in the form:
\begin{equation}
O_n (z , {\bar{z}} )= e^{i p_n X_{0} } + e^{-ip_n X_0 }  , \quad p_n = i n
{{ \pi}\over{\beta}} \left ( {{\alpha^{\prime} }\over{ 2}} \right )^{1/2} ,  \quad
n= 1, \cdots \infty \quad .
\label{mom}
\end{equation}
It is clear that a coherent superposition of the $O_n$ will be
localized at the fixed points of the orbifold, $X_{0}$$=$$0$,
$\beta$. At temperature $T_{\rm min}$, all of the $O_n$ with
$|n|$$\ge$$2$ are irrelevant, while $O_1$ is marginally relevant.
Thus, $O_1$ drives the RG flow, in the direction of a decrease in
$g$ and, consequently, an increase in $\beta$. Without computing
$g$ we cannot tell whether the direction of the flow is towards
the thermal vacuum at $T_{\rm min}$, or away from it, and so we
must proceed with an explicit calculation. Consider the expression
for the free energy given in \cite{fermionic}. In terms of the
dimensionless variable, $x^2$$=$$2\beta^2/\pi^2 \alpha^{\prime}$,
we have,
\begin{eqnarray}
{\cal F} =&& - {{1}\over{\beta}}
\int_{ F} {{d^2 \tau}\over{4\tau_2^2}}
  (2 \pi \tau_2 )^{-4} |\eta(\tau)|^{-22} \times
\cr
 &&  {{1}\over{2}} {{1}\over{\eta{\bar{\eta}}}} \left [ \sum_{w,n \in {\rm Z} }
   q^{{{1}\over{2}} ({{n }\over{x}} + {{wx}\over{2}} )^2 }
     {\bar q}^{{{1}\over{2}} ({{n}\over{x}} - {{wx}\over{2}} )^2 }
 \right ]
\{ (|\Theta_3 |^8
  + |\Theta_4|^8  + |\Theta_2|^8 ) \cr
  \quad&& + e^{ \pi i (n + w ) }
 [ (\Theta_2^4 {\bar{\Theta}}^4_4 + \Theta_4^4 {\bar{\Theta}}_2^4)
  - (\Theta_3^4 {\bar{\Theta}}_4^4 + \Theta_4^4 {\bar{\Theta}}_3^4
   + \Theta_3^4 {\bar{\Theta}}_2^4 + \Theta_2^4 {\bar{\Theta}}_3^4 )
   ]\}\label{eq:bosod}
\end{eqnarray}
The $n$$=$$w$$=$$0$ sector of this expression is the free energy of the supersymmetric
type II string, identically zero as a consequence of the equality of target space fermions
and bosons at each mass level. The tachyonic state in the $n$$=$$w$$=$$0$ sector has
been projected out. The $n$$=$$1$, $w$$=$$0$ state is marginally relevant at $T_{\rm min}$,
there are no relevant operators in the conformal field theory, and the free energy of the
thermal vacuum of the type II string at this temperature is normalizable. At temperature
$T_2$$=$$1/8 \pi^2 \alpha^{\prime}$, the operator $O_2$ turns marginally relevant, but $O_1$ is
still the most relevant operator, therefore driving the RG flow. As we tune the
temperature to still lower
values, this pattern continues with successive thermal momentum modes turning marginally
relevant, but the most relevant operator continues to be $O_1$. Thus, it is clear that
$O_1$ drives the RG flow all the way from, or to, the noncompact, zero temperature limit.
In this limit, the system approaches a new, ultraviolet or infrared, fixed point and the
coherent superposition of thermal modes decouples from the worldsheet sigma model.
To distinguish ultraviolet from infrared, we will now carry out the computation of
${\rm ln} g$.

\vskip 0.1in
Consider the factorization limit of the torus, dominated by contributions from the lowest
lying closed string modes or, conversely, the high energy asymptotic behavior since the
expression above is invariant under the modular transformation $\tau$ $\to$ $-1/\tau$. In
the factorization limit, we can express the theta functions in terms of their
$q$-expansions, and restrict to the leading term in the integrand. For the thermal vacuum
at temperature $T_{\rm min}$, we obtain the result:
\begin{equation}
{\rm ln } g_{\rm T_{min} } =  - \beta \lim_{\tau_2 \to \infty}
{\cal F} (\beta) = \int_{ F} {{d^2 \tau}\over{4\tau_2^2}}
  (2 \pi \tau_2 )^{-4}  \left ( 2 + (16 -16) + O(e^{-2\pi\tau_2}) \right ),
\label{eq:betab}
\end{equation}
where the factor of unity comes from the marginally relevant thermal perturbation.
Notice that a phase factor of $(-1)$ in the $n$$=$$1$, $w$$=$$0$ sector
results in a nonvanishing contribution from the terms in $\Theta_3$ and $\Theta_4$,
as opposed to the cancellation in the $n$$=$$w$$=$$0$ sector due to a $(+1)$ phase.
This is the phase factor leading to the absence of a tachyon in the usual
superstring.

\vskip 0.1in
Next, we compare with the value of ${\rm ln} g$ evaluated at the noncompact
zero temperature limit \cite{fermionic}:
\begin{eqnarray}
{\rm ln} g_{\rm 0 } =&&
 \lim_{\tau_2 \to \infty}
\int_{ F} {{d^2 \tau}\over{4\tau_2^2}}
  (2 \pi \tau_2 )^{-4} |\eta(\tau)|^{-24}
  \left [ (\Theta_3^4 - \Theta_4^4 - \Theta_2^4)
   ({\bar{\Theta}}_3^4 - {\bar{\Theta}}_4^4 - {\bar{\Theta}}_2^4 )
\right ] \cr
 =&& \lim_{\tau_2 \to \infty}
\int_{ F} {{d^2 \tau}\over{4\tau_2^2}}
  (2 \pi \tau_2 )^{-4} \left ( 16 - 16 + O(e^{-2\pi \tau_2} ) \right )
\quad .
\label{gss}
\end{eqnarray}
This quantity vanishes as a consequence of target space supersymmetry, however,
the separate contributions from bosonic and fermionic are nonvanishing.
If we restrict ourselves to the bosonic degrees of freedom in the
worldsheet sigma model as suggested by HKMM, we find that ${\rm ln} g$ is a
finite number at either fixed point. It is expressed in terms of an unambiguously
normalized integral over the fundamental domain of the torus. Most importantly,
the numerical coefficient for the bosonic degrees of freedom {\em decreases} from
$18$ to $16$. We conclude that the thermal type IIA vacuum is unstable
against thermal perturbations: the infrared fixed point is at $T$$=$$0$, and
the RG flow is in a direction {\em towards} the supersymmetric vacuum.

\vskip 0.1in
If we now repeat the calculation of ${\rm ln} g$ at any intermediate temperature
$T^2_n$$=$$1/2n^2\pi^2 \alpha^{\prime}$ where the $n$th momentum mode is turning
marginally relevant, we obtain a result identical to Eq.\ (\ref{eq:betab}) for
the {\em normalizable} states in the worldsheet sigma model. In addition, the
sigma model has non-normalizable states which correspond to target space tachyons.
Their contribution to the factorization limit of the path integral is divergent
and, following the suggestion of \cite{hkmm}, we should excise such states from
the computation of ${\rm ln} g$ in order to obtain a physically sensible answer.
It is nevertheless interesting that Eq.\ (\ref{eq:bosod}) gives an explicit
result at intermediate points along the RG flow for {\em both} nonrmalizable and
non-normalizable contributions to the free energy. Separating these contributions,
as suggested by statement (iii) of AL, we can express the free energy in the form:
\begin{equation}
{\cal F} (\beta) = f_{\rm ST} - {{1}\over{\beta}} {\rm ln} g
\quad ,
\label{eq:nonuni}
\end{equation}
where ${\rm ln} g$ is universal, and $ f_{\rm ST}$ is a non-universal contribution
to the target space free energy. Notice that the IR divergence on the
worldsheet may be freely
interpreted instead as a target space UV divergence in the free energy. We remind
the reader that, while ${\cal W}$
is unambiguously normalized, the free energy of a system can in general only
be computed up to an overall scaling of inverse temperature: it is the {\em ratio}
$F/T$ which has possible numerical relevance. This statement holds for both quantum
field theory and string theory. At intermediate points along the RG flow we have
the following formal expression for the nonuniversal contribution to the free
energy:
\begin{equation}
- \beta f_{\rm ST} =
\int_{ F} {{d^2 \tau}\over{4\tau_2^2}}
  (2 \pi \tau_2 )^{-4}
 {{1}\over{2}} e^{2\pi \tau_2}
 \sum_{\Delta_{nw} < 1, n,w \in {\rm Z} }
   e^{- 2\pi \tau_2 ({{n^2 }\over{x^2}} + {{w^2x^2}\over{4}})} e^{2\pi i nw \tau_1}  \quad ,
\label{eq:tachy}
\end{equation}
where $T$$<$$T_{\rm min}$, and $\Delta_{nw}$ is the conformal
dimension of the $(n,w)$th mode. Any attempt to extract a finite
answer from this divergent integral will introduce regulator
dependence in the expression for the free energy. In its current
form the expression for $f_{\rm ST}$ preserves all of the
worldsheet super-Diff$\times$super-Weyl gauge symmetries.

\vskip 0.1in
Our results are clear confirmation of a conjecture due to Adams, Polchinski, and
Silverstein \cite{aps}, namely, that nonsupersymmetric type II orbifolds generically
decay under worldsheet RG flows to the supersymmetric flat space type II vacuum.
It is interesting to ask what happens at the other marginally relevant fixed
point of this system, the thermal type IIA vacuum at temperature
$T^2_{\rm max}$$=$$2/\pi^2 \alpha^{\prime}$ \cite{fermionic}. By thermal duality,
we can predict that the direction of RG flow is away from the thermal vacuum at
$T_{\rm max}$, and in the direction of increasing temperature. Since the type IIA
and type IIB thermal vacua are related by a thermal duality, this behavior for the
type IIA string is simply an equivalent way of expressing the flow from
$T_{\rm min}$ to $T=0$ for the dual type IIB string.

\vskip 0.1in It is of interest to examine the behavior of the
marginally stable thermal vacuum of the heterotic string theory
within this same framework. In \cite{fermionic}, we showed that
the thermal vacuum of the heterotic string is tachyon free at all
temperatures in the presence of a temperature-dependent Wilson
line. All of the operators carrying thermal momentum correspond to
irrelevant flows of the worldsheet RG. Thus, the possibility of
marginal relevance does not arise and inverse temperature does not
correspond to a coupling in the space of RG flows for the
worldsheet sigma model. We emphasize that this is the result in
perturbation theory at fixed string coupling. Since the thermal
heterotic vacuum has a non-vanishing one-loop cosmological
constant, it is likely to exhibit a two-loop contribution to the
dilaton tadpole. Thus, it is still possible that in the extended
space of RG flows containing both the dilaton vev, or string
coupling, and the inverse temperature, one can observe a
nontrivial flow of the worldsheet RG. We leave a discussion of
this subtle case for future work. In \cite{fermionic}, we have
conjectured that the heterotic thermal vacuum flows to strong
coupling, and that this strongly coupled heterotic vacuum is the
weakly coupled type I string vacuum described in \cite{fermionic}.
It would be most illuminating to understand the nature of such RG
flows.

\vskip 0.1in
Finally, we come to a discussion of the thermal vacuum of the type I string.
The free energy is given by summing over worldsheets with the topology of a
cylinder, Mobius strip, and Klein bottle. The result is \cite{fermionic}:
\begin{eqnarray}
F = && \beta^{-1}
\int_0^{\infty} {{dt}\over{8t}}
  {{(2\pi t )^{-9/2}}\over{
  \eta(it)^{8}}}
[ ~  2^{-8} N^2 ( Z_{\rm [0]} - e^{\i \pi n } Z_{\rm [1]})
 \cr
\quad && \quad\quad
+   Z_{\rm [0]} - e^{\i \pi n } Z_{\rm [1]}
 ~ - ~ 2^{-3} N  Z_{\rm  [0]} \cr
\quad&& \quad\quad\quad
- ~ 2^{-3} N e^{\i \pi n } Z_{\rm [1]} ~ ] \times
\sum_{n \in {\rm Z}; w=0,1 } q^{\alpha^{\prime}\pi^2 n^2 /\beta^2 +
   w^2 \beta^2/ 4 \pi^2 \alpha^{\prime} }
\label{eq:freeIp}
\end{eqnarray}
The Matsubara frequency spectrum is given by the eigenvalues for timelike
momentum: $p_n$$=$$2n\pi/\beta$, with $n$$\in$${\rm Z}$. Notice that $w$
only takes the values $0$, $1$. The functions $Z_{[0]}$, $Z_{[1]}$ are
defined as follows \cite{polbook}:
\begin{eqnarray}
Z_{\rm [0]} =&& ({{\Theta_{00}(it;0)}\over{\eta(it)}})^4 -
({{\Theta_{10}(it;0)}\over{\eta(it)}})^4
\cr
Z_{\rm [1]} =&& ({{\Theta_{01}(it;0)}\over{\eta(it)}})^4 -
({{\Theta_{11}(it;0)}\over{\eta(it)}})^4
\label{eq:relats}
\end{eqnarray}
where the (00), (10), (01), and (11), denote, respectively, (NS-NS), (R-NS),
(NS-R), and (R-R), boundary conditions on worldsheet fermions in the closed string sector.
The boundary conditions in the open string sectors can be read off using worldsheet
duality \cite{polbook}. As in the supersymmetric zero temperature limit, the
independent cancellation of both the dilaton and the Ramond-Ramond ten-form tadpole
requires that $N$$=$$2^4$. At finite temperature, this cancellation mechanism also
serves to project out potential tachyonic thermal perturbations in the (00) and
(01) sectors from the free energy. We emphasize that the projection {\em requires} the
addition of crosscap contributions to the disc amplitudes obtained in the factorization
limit, and with the weights given in Eq.\ (\ref{eq:freeIp}).

\vskip 0.1in
It is interesting to ask what extension of the $g$-theorem is appropriate in
the case of an unoriented open and closed string theory. If we include crosscap
contributions to ${\rm ln} g$ in addition to the disc, ${\rm ln} g$ vanishes identically,
even when restricted to target space bosonic degrees of freedom. Nevertheless, we
feel this is the correct answer since there are no marginally relevant flows
of the worldsheet RG in this case and, consequently, no thermal instabilities.
We emphasize that ${\rm ln} g$ vanishes both at finite temperature, and in the
noncompact supersymmetric zero temperature limit. In addition, since the oneloop
cosmological constant vanishes there is no signal of a dilaton tadpole. Thus, the
thermal vacuum appears stable even in the extended space of RG flows including the
dilaton vev.
In \cite{fermionic}, we showed that the type I string theory can also exist in a distinct
long string phase at high temperature, and that this phase is fully accessible even in the
limit of zero string coupling. The absence of thermal instabilities establishes the existence
of a vacuum in true thermal equilibrium in type I string theory, and in either
of two distinct phases.
Our results demonstrate, on the other hand, that the
type II strings do not achieve a state of time-independent thermal equilibrium, at
least not in the absence of Ramond-Ramond backgrounds \cite{fermionic}.
As has been suggested in different contexts \cite{aps,hkmm,martinec,ghms},
the presence of marginally relevant flows of the worldsheet RG indicates that the
vacuum in question is unstable against decay, evolving in real time towards a more stable
vacuum of string theory.

\vskip 0.1in We will end with a conjecture. As mentioned earlier,
there is a precise correspondence between the open string tree
level graphs and the tree-level spacetime effective action and
this fact has been the basis for subsequent developments in open
string field theory. HKMM have already noted that this implies a
relation between $\Gamma_{\rm ST}$ and $g$, the vacuum degeneracy
of normalizable bosonic degrees of freedom: $g $$=$$\half
\Gamma^2_{\rm ST}$ \cite{hkmm}. However, the factorization limit
of the closed string amplitude is also normalizable, offering, as
we have shown, a direct route to the computation of $g$ for closed
string theories. This suggests the following modified form of the
perturbative RG flow equations:
\begin{equation}
e^{\half {\rm ln} g} {{\partial g}\over{\partial \lambda_i}} = \beta_i(\lambda)G_{ij} , \quad
e^{\half {\rm ln} g } {{\partial g}\over{\partial \mu}} = - \beta_i \beta_j G_{ij}  \quad .
\label{eq:betag}
\end{equation}
These equations can be used to infer the form of the perturbative beta functions, $\beta_i$,
in terms of the Zamolodchikov metric, $G_{ij}$, in the space of couplings, $\{ \lambda_i \}$,
the vacuum degeneracy, $g$, and the worldsheet renormalization scale, $\mu$. Each of these
quantities has been given a precise worldsheet definition, and the flow equations hold for
both closed, and open and closed, string theories. It would be most interesting to explore
the nature of these equations in detail.

\noindent {\bf ACKNOWLEDGMENTS:} I would like to thank G.\ Moore
for an email correspondence.

\vskip 0.5in
\noindent{\bf Note Added (Sep 2005):} Many of the points made in this
paper are either extraneous, or incorrect in the details, although the broad
conclusion stands. Namely, that it is important to have a well-defined 
criterion such as the g--theorem, or suitable adaptation, that determines 
the direction of worldsheet RG flows describing
the vacuum configuration space of string theories. The pedagogical review 
of the Affleck and Ludwig's g-theorem, distinguishing
its application in field theory and string theory, is of value in this context. 
I refer the reader 
to hep-th/0506143 for discussion of the string canonical ensembles.

\end{document}